\documentclass[10pt,letterpaper,twocolumn]{article}

\usepackage{graphicx}
\usepackage{ol2}
\begin{document}
\twocolumn[

\title{Precise multi-emitter localization method for fast super-resolution imaging}

\author{Yuto Ashida,$^{1}$ Masahito Ueda$^{1,2}$}

\address{
$^1$Department of Physics, University of Tokyo, 7-3-1 Hongo, Bunkyo-ku, Tokyo
113-0033, Japan.\\
$^2$Center for Emergent Matter Science (CEMS), RIKEN, Wako, Saitama 351-0198, Japan.
}

\begin{abstract}
We present a method that can simultaneously locate positions of overlapped multi-emitters at the theoretical-limit precision. We derive a set of simple equations whose solution gives the maximum likelihood estimator of multi-emitter positions. We compare the performance of our simultaneous localization analysis with the conventional single-molecule analysis for simulated images and show that our method can improve the time-resolution of superresolution microscopy an order of magnitude. In particular, we derive the information-theoretical bound on time resolution of localization-based superresolution microscopy and demonstrate that the bound can be closely attained by our analysis. 
\end{abstract}


]

Precisely and accurately locating point objects is a long-standing common thread in science. Recent realizations of super-resolved imaging of single molecules \cite{TAK00,MGLG05,EB06} have revolutionized our view of quasi-static nanostructures {\it in}-{\it vivo}. In particular, a wide-field approach based on localizing individual fluorophores has emerged as a versatile method \cite{EB95,HST06,RMJ06,AP10}. The single-molecule localization microscopy (SMLM) works under conditions in which fluorescent molecules are activated at very low density so that no more than a single molecule within any diffraction-limited region emits photons simultaneously \cite{GP10}. A set of single-molecule positions can then be precisely determined beyond the diffraction limit by fitting each image of molecules using a single point spread function. Nevertheless, the slow temporal resolution of super-resolved imaging severely restricts the utility to the study of live-cell phenomena. This is because the analysis discards the information from crowded molecules with overlapping images through filtering and, typically, SMLM requires accumulating thousands of frames to generate a super-resolution image \cite{BH10,LS10}. A substantial reduction of the imaging time will significantly expand the horizon of super-resolution techniques and enable to observe fast, nanoscale dynamics {\it in}-{\it vivo}. 

Recently, there have been remarkable progresses in improving the temporal resolution of super-resolution microscopy by developing multi-emitter localization algorithms \cite{SJH11,TQ11,SC12,FH11,YW12,FH13,AS14}. In particular, Ref. \cite{FH13} has utilized sCMOS camera to achieve an impressively high imaging speed (32 frame/s). Also, there exists now commercially available super-resolution microscopy for live-cell imaging from Nikon. Yet, the theoretical limit on the time resolution has remained elsusive and hence, to maximize the potential of the technique, a multi-emitter localization analysis that allows us to attain the theoretical bound needs to be developed.

Here we develop a multi-emitter localization analysis that can attain at nearly the theoretical-limit speed of localization-based superresolution microscopy. To begin with, let us discuss the theoretical limit on the temporal resolution of super-resolved imaging. The image acquisition time $T_{\rm img}$ of localization-based superresolution microscopy is determined by a number of frames and an exposure time of each frame. We define the fidelity $F$ of super-resolved imaging as the fraction of imaged molecules i.e., activated molecules at least once during the entire process of imaging. Then the required number of frames to ensure the fidelity is given by  $-\ln(1-F)\rho_{\rm obj}/\rho_{\rm img}$, where $\rho_{\rm obj}$ ($\rho_{\rm img}$) is the object (image) molecule density \cite{ARS09}. An exposure time of a single frame is determined by the required number of photons $N_{\rm photon}$ to achieve the desired resolution $\delta$ divided by the collection rate $\rho_{\rm img}S\Gamma$, where $S$ is the area of the region of interest and $\Gamma$ is the collection rate of photons per molecule. While the switching rate of molecules is assumed to be optimized with the exposure time, this should not be considered to be realizable in a single experimental setup; here we are interested in the fundamental limit among all experimental situations. Information theory dictates that $N_{\rm photon}$ be bounded by the Fisher information matrix \cite{TMC06}. To obtain the net information gain, we assume the point spread function (PSF) of a single molecule as a 2D Gaussian with a standard deviation $\sigma$ and calculate the precision limit by assuming an arrayed configuration of molecules with density $\rho_{\rm img}$. Then, we evaluate the Fisher information matrix \cite{TMC06} and its inverse numerically. We define the diagonal element of the inverse matrix in multi-molecule-position basis as $[\mathcal{I}^{-1}]_{DD}$. Consequntly, $N_{\rm photon}$ is bounded as
\begin{equation}
N_{\rm photon}(\rho_{\rm img},\delta,\sigma)\geq\frac{1}{\delta^{2}}[\mathcal{I}^{-1}]_{DD}\equiv\frac{1}{\delta^{2}}\rho_{\rm img}S\Delta^{2}(\rho_{\rm img},\sigma),
\end{equation}
where we define the normalized precision limit $\Delta^{2}$ by the last equality. While $\Delta^{2}$ has a simple relation $\Delta^2=\sigma^2$ when molecules are sufficiently sparse, its value can be obtained only numerically in a high-density regime.
The information-theoretic limit on the image acquisition time is now obtained by
\begin{equation}\label{time}
T_{\rm img}\geq\ln\Bigl(\frac{1}{1-F}\Bigr)\frac{\rho_{\rm obj}}{\rho_{\rm img}^{*}}\frac{\Delta^{2}(\rho_{\rm img}^{*},\sigma)}{\delta^{2}\Gamma},
\end{equation}
where $\rho_{\rm img}^{*}$ is the optimized image density so that the lower-bound of the image acquisition time is minimized. 

\begin{figure}[t]
\begin{center}
\fbox{\includegraphics[width=\linewidth]{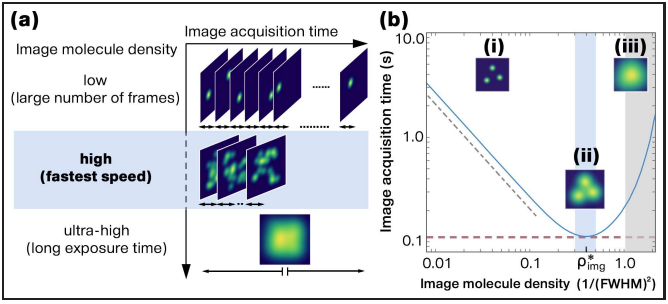}}
\end{center}
\caption{(Color online) (a) Schematic figure about temporal resolution of localization-based superresolution microscopy. There is an optimal image molecule density that allows the fastest image acquisition. (b) Theoretical limit on the image acquisition time plotted against the image molecule density, calculated for $\delta=10$nm, fidelity $F=0.9$, $\Gamma=2.3\times10^{4}$ 1/s, and $\sigma=82.5$nm. The fastest imaging can be achieved in the blue-shaded region, which sets the theoretical limit indicated by the red-dashed line.}
\label{fig1}
\end{figure}

Figure \ref{fig1}a summarizes the theoretical limit on the time resolution of super-resolution microscopy for varying image density. In a low-density region, where the activated molecules are sparsely distributed so that the interference patterns rarely overlap (see also the inset figure (i) in Fig. \ref{fig1}b), the amount of information carried by a single photon is nearly constant. Thus, increasing an image density directly reduces the acquisition time for a super-resolution image which scales as $\propto \rho_{\rm img}^{-1}$ (indicated by the black-dashed line in Fig. \ref{fig1}b). However, once the interference patterns significantly overlap and the peaks of molecules cannot be resolved anymore (see the inset figure (iii)), the information gain per photon dwindles rapidly. Consequently, the required number of photons rapidly increases, resulting in a sharp increase in the image acquisition time. Hence, the fastest time resolution is achieved at a high image density between the above two situations, where the information acquisition rate is maximal (see  the inset figure (ii)). Such density regime is indicated by the blue-shaded region in Fig. \ref{fig1}b. To achieve the theoretical-limit speed, the crucial step is to develop a multi-emitter localization analysis that can faithfully work under such high density region. As detailed later, our method can closely attain such limit. While the theoretical curve in Fig. \ref{fig1}b makes sense irrespective of the performance of a specific algorithm, we indicate, for convenience, the region where our multi-emitter localization method fails to work by the gray-shaded region. 

We now describe our simultaneous localization analysis. In a previous work, we show that tracking progressive collapse of many-body wavefunction enables a diffraction-unlimited position measurement of ultracold atoms \cite{YA15} in an optical lattice. We here generalize this approach to classical objects, such as fluorescent molecules, by treating the estimated position distribution as a counterpart of the quantum-mechanical wavefunction. We model the effective point spread function (PSF) of a single molecule by a 2D Gaussian $P[\mathbf{r}|\mathbf{R}]=\exp(-|\mathbf{r}-\mathbf{R}|^{2}/(2\sigma^{2}))/(2\pi\sigma^2)$, where $\sigma=0.21\lambda/{\rm NA}$ \cite{BZ07} is the standard deviation with NA being the numerical aperture, and we denote the position of the molecule as $\mathbf{R}$ and that of photodetections as $\mathbf{r}$.  The interference pattern of multi-emitter is constructed from an incoherent sum of these point spread functions: $P[\mathbf{r}|\{\mathbf{R}\}]=(1-\epsilon)/N\sum_{m=1}^{N} P[\mathbf{r}|\mathbf{R}_{m}]+\epsilon/S$, where we denote $\{\mathbf{R}\}\equiv\{\mathbf{R}_{1},\mathbf{R}_{2},\ldots,\mathbf{R}_{N}\}$ as a set of $N$ molecule positions, $\epsilon$ is the fraction of the background noise, which can be related to the signal-to-noise ratio (SNR) as $\epsilon=1/(1+\rm{SNR})$, and $S$ is the area of the region of interest. We formulate the imaging as a stochastic process in which spatial locations of photodetections are randomly generated according to the interference pattern. 

Let us assume that $M$ photons are detected at $\mathbf{r}_{1},\mathbf{r}_{2},\ldots,\mathbf{r}_{M}$. The conditional probability distribution of a set of $N_{\rm est}$ molecule estimators $\{\mathbf{R}_{\rm est}\}\equiv\{\mathbf{R}_{1},\mathbf{R}_{2},\ldots,\mathbf{R}_{N_{\rm est}}\}$ is given by the Bayesian inference:
$P[\mathbf{\{R\}_{\rm est}}|\mathbf{r}_{1},\mathbf{r}_{2},\ldots,\mathbf{r}_{N_{\rm photon}}]\propto\prod_{i=1}^{M}P[\mathbf{r}_{i}|\{\mathbf{R}\}_{\rm est}]P_{0}[\{\mathbf{R}\}_{\rm est}]$, where $P_{0}[\{\mathbf{R}\}_{\rm est}]$ represents a prior distribution of the molecule distribution. Since we assume no prior knowledge about the configuration of molecules, the initial distribution $P_{0}$ is chosen to be a uniform distribution. In our formulation, the problem of identifying the most probable set of molecule positions is equivalent to maximizing the conditional probability distribution with respect to possible multiple-molecule configurations $\{\mathbf{R}_{\rm est}\}$. 

Remarkably, for Gaussian point spread functions, we can show by analytical calculations that the problem of finding the most probable set of estimators is substantially simplified to solving the following $N_{\rm est}$ self-consistent equations:

\begin{equation}\label{self}
\mathbf{R}_{m}=\frac{\sum_{i=1}^{M}\mathbf{r}_{i}g_{m}\bigl(\mathbf{r}_{i};\{\mathbf{R}\}_{\rm est}\bigr)}{\sum_{i=1}^{M}g_{m}\bigl(\mathbf{r}_{i};\{\mathbf{R}\}_{\rm est}\bigr)},
\end{equation}
where $m=1,2,\ldots,N_{\rm est}$ is the label of each estimator. We then introduce the weight-function $g_{m}$ by
\begin{equation}\label{gm}
g_{m}\bigl(\mathbf{r};\{\mathbf{R}\}_{\rm est}\bigr)\equiv\frac{\exp\Bigl(-\frac{|\mathbf{r}-\mathbf{R}_{m}|^{2}}{2\sigma^{2}}\Bigr)}{\gamma+\sum_{k=1}^{N_{\rm est}}\exp\Bigl(-\frac{|\mathbf{r}-\mathbf{R}_{k}|^{2}}{2\sigma^{2}}\Bigr)}.
\end{equation}
Here we define the term $\gamma\equiv 2\pi\sigma^{2}N_{\rm est}\epsilon/\bigl(S(1-\epsilon)\bigr)$ which describes the contribution from the background noise. This type of equations can be efficiently solved by using standard numerical methods \cite{RLB04}. We note that the estimators coincide with the maximum likelihood estimator and hence, the theoretical-limit precision, i.e., the Cram\'er-Rao bound, can be asymptotically attained. We can easily generalize the above discussions to pixelated measurements in which the number of photodetections $N_{i,j}$ at each pixel $\langle i,j\rangle$ constitutes the sufficient statistic. The result is the following set of $N_{\rm est}$ self-consistent equations for estimators:
\begin{equation}\label{selfpix}
\mathbf{R}_{m}=\frac{\sum_{\langle i,j\rangle}\mathbf{r}_{i,j}g_{ij;m}\bigl(\{\mathbf{R}\}_{\rm est}\bigr)}{\sum_{\langle i,j\rangle}g_{ij;m}\bigl(\{\mathbf{R}\}_{\rm est}\bigr)},
\end{equation}
where $\mathbf{r}_{i,j}$ is the pixel position and $g_{ij;m}\equiv N_{ij}g_{m}\bigl(\mathbf{r}_{i,j};\{\mathbf{R}\}_{\rm est}\bigr)$. Note that such simple equations have not been derived in other approaches of multi-emitter localization \cite{SJH11,TQ11,SC12,FH11}. 

The above formulation can be also extended to the case with a nonuniform background noise. Let $\epsilon(\mathbf{r})$ be the fraction of the background noise at position $\mathbf{r}$. By replacing the term $\gamma$ in Eq. (\ref{gm}) with
$
\gamma(\mathbf{r})\equiv(2\pi\sigma^{2}N\epsilon({\mathbf{r}}))/(1-\int_{S}\epsilon(\mathbf{r}'){\rm d}\mathbf{r}'),
$
we can use the self-consistent equations in Eq. (\ref{self}) to obtain the most probable estimators for molecule positions with nonuniform background. 
If we utilize the astigmatism \cite{HL07}, a generalization to 3D imaging is also possible. The point spread function is described by an asymmetric Gaussian function whose widths are given by $\sigma_{X,Y}(Z)=\sigma_{0}\sqrt{1+(Z\mp\eta)^2/d^2}$, where $\eta$ is an axial astigmatism, $\sigma_0$ is the focus width, $d$ is the focus depth. The axial position $Z_{m}$ of the molecule $m$ is obtained as the solution of the self-consistent equation,
\begin{equation}
Z_{m}=\eta\cdot\frac{\sum_{i=1}^{M}(g_{i;m}^X-g_{i;m}^Y)}{\sum_{i=1}^{M}(g_{i;m}^X+g_{i;m}^Y)}.
\end{equation}
We introduce the weight functions $g_{i;m}^{\alpha}=f_{i;m}^{\alpha} e^{-h_{i;m}}/(\gamma+\sum_{k=1}^{N_{\rm est}}e^{-h_{k;m}}\sigma^2_{0}/(\sigma_{X}(Z_{k})\sigma_{Y}(Z_{k})))$, where $f_{i;m}^{\alpha}\equiv((a_{i}-\alpha_{m})^2-\sigma_{\alpha}^2(Z_{m}))/\sigma_{\alpha}^4(Z_{m})$, $h_{i;m}\equiv(x_{i}-X_{m})^2/(2\sigma_{X}^2(Z_{m}))+(y_{i}-Y_{m})^2/(2\sigma^2_{Y}(Z_{m}))$, $\alpha=X,Y$, and $a=x,y$.
These generalizations can be applied jointly to deal with realistic situations.


To solve the self-consistent equations (\ref{selfpix}), we apply a high-order iterative method known as Steffensen's method \cite{RLB04}, which allows quadratic convergence. In contrast to an ordinary iterative method, this method does not need to calculate derivatives and, in our problem, can efficiently perform calculations. To avoid an unwanted divergence and make a robust convergence, when a temporal value of estimators becomes unreasonably high, the simple successive substitution is  concomitantly used. The iteration is terminated when the calculation converges or after 100 iterations. 

Finding the global maximum of the conditional probability distribution is an essential step for the simultaneous localization. To do so, we combine the pre-estimation and the optimization based on the information measure as follows. First, we pre-estimate and localize molecules based on the single-molecule analysis with the well-established rejection algorithm in which the local maximum is identified by setting a suitable threshold \cite{SJH11}. Then, the image is fitted with a single point spread function and the position of a molecule is estimated. If the result of the fitting significantly deviates from the position of the local maximum, the image is judged as constructed from multiple molecules and the estimated position is discarded. The resulting set of $N_{\rm ini}$ localized positions constitutes the first set of initial positions for iterations.

Second, for each assumed number of molecules $N_{\rm est}$, the residual $N_{\rm est}-N_{\rm ini}$ initial positions are randomly generated according to the observed probability distribution of photodetections. This enables the well-estimated initialization of estimators in successive iterative calculations. A large number of different sets of initial positions are generated. Then, iterative calculations to solve the self-consistent equations (\ref{selfpix}) are performed by starting from each set  (typically, preparing about one thousand different sets is sufficient to find the global solution at the high-molecule density). The iteration result that minimizes the Kullback-Leibler divergence between the observed distribution and the distribution reconstructed from the estimated positions is chosen as the most probable set of molecule positions within the sector of $N_{\rm est}$-molecule configuration space. As a typical processing speed, the analysis for 60$\times$60 pixels with 100 iterations can be finished by the CPU calculation with C code program within about 3.5 s.
The performance of the calculation can be enhanced by increasing both the trial number of preparing the initial configurations and the number of iterations.

\begin{figure}[t]
\begin{center}
\fbox{\includegraphics[width=\linewidth]{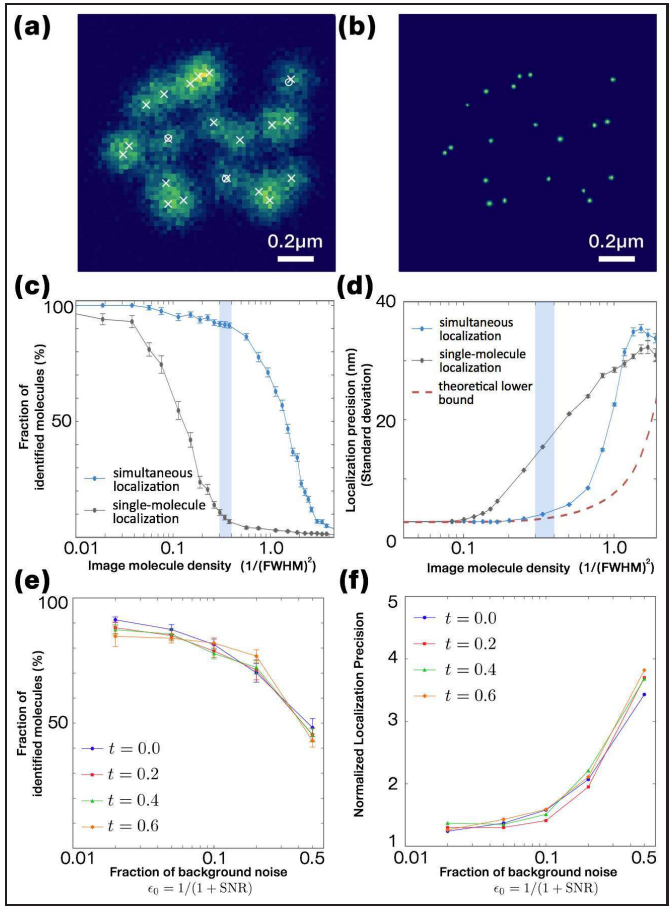}}
\end{center}
\caption{\label{fig2}(Color online) (a) A simulated image of fluorescent molecules with uniform background noise ($\rho_{\rm img}=16 /\mu$m$^2$, SNR=50, pixel size=28 nm). The crosses indicate the true positions of particles. SMLM only identifies emitters indicated by circles. (b) The result of our method. (c,d) The fraction of identified molecules and the localization precision, calculated for $10^3$ photons per molecule, SNR=50, and $10^4$ simulated images. In (d), the theoretical lower bound is shown by the red-dashed curve. (e,f) The performance of our method in the low SNR and in the presence of nonuniform noise, calculated for $\rho_{\rm img}=10 /\mu$m$^2$, $10^3$ photons per molecule, and $10^2$ images. 
}
\end{figure}

We perform the above procedures of estimating the most probable set of molecule positions for various molecule number $N_{\rm est}\geq N_{\rm ini}$. For each iteration result of the $N_{\rm est}$-molecule configurational space, we calculate the expected probability distribution $P_{N_{\rm est}}$ of photodetections and compare it with the observed probability distribution $P_{\rm data}$ by employing the Kullback-Leibler divergence $D[P_{\rm data}|P_{N_{\rm est}}]$. The final set of the most probable molecule positions is determined by minimizing the Kullback-Leibler divergence with respect to $N_{\rm est}$.

To demonstrate how superior our simultaneous localization approach is to the conventional SMLM, we show a typical result in Fig. \ref{fig2}a, b. While SMLM can only identify the well-isolated molecules as indicated by the circles in Fig. \ref{fig2}a, our method identifies all molecules at the theoretical-limit precision as shown in Fig. \ref{fig2}b. Hence, our analysis allows an accurate and precise localization of multi-emitters despite a substantial overlap of images. We note that our value of the pixel size does not compromise the superiority of our method since an expected improvement of the localization precision  with respect to a larger pixel size (100nm) is only 5\% \cite{TRE02}.

To achieve the fundamental limit of time resolution in Eq. (\ref{time}), the crucial fact is that our simultaneous localization significantly outperforms the single-molecule analysis at high density region, where super-resolved imaging can be performed with ultimate time resolution. We demonstrate this by applying the simultaneous localization and SMLM analyses to different simulated images repeatedly. The simultaneous localization analysis attains more than a tenfold improvement of the fraction of identified molecules (the so-called ``recall" \cite{SW11}) with respect to the conventional single-molecule analysis (Fig. \ref{fig2}c). Also, this is a fourfold improvement compared with the reported performance of DAOSTORM at the same signal-to-noise ratio (SNR) \cite{SJH11}. Note that our method achieves some ninety percent accuracy in the blue-shaded region enabling the fastest super-resolved imaging. The simultaneous localization also achieves, in the region of the ultimate time resolution, the theoretical-limit precision (Fig. \ref{fig2}d). In particular, our method can closely attain the precision limit up to $\rho_{\rm img}\simeq 13 /\mu$m$^2$ (with $\sigma=82.5$nm), which shows again about a fourfold improvement compared with DAOSTORM \cite{SJH11}. At an ultra-high density region where the density becomes larger than the optimal region and the images completely overlap (indicated by the gray-shaded region in Fig. \ref{fig1}b), our analysis fails to localize molecule positions. This is  because the information gain carried by a photon almost vanishes and an extremely high signal-to-noise ratio is required to achieve the desired precision. 

Our method also works in the low SNR and in the non-uniform background noise. To demonstrate this, we perform numerical simulations by assuming the noise configuration $\epsilon(\mathbf{r})=\epsilon_{0}(1+te^{-\mathbf{r}^2/(2\sigma'^2)})/\mathcal{N}$, where $\epsilon_{0}=1/(1+{\rm SNR})$, $\mathcal{N}\equiv\int_S d\mathbf{r}'(1+te^{-\mathbf{r}'^{2}/(2\sigma'^2)})$, $\sigma'=4\sigma$ is the width of the nonuniformity. The origin of the coordinates represents the center of the region of interest. Figures \ref{fig2}e-f show the fraction of identified molecules and the localization precision normalized by the theoretical limit $\Delta$ in the high-molecule density $\rho_{\rm img}=10 /\mu$m$^2$ (with $\sigma=82.5$nm). In particular, in the high background noise condition (SNR$\simeq$ 8), about 80\% recall can be achieved by our method, and this makes an improvement of the image molecule density by a factor of about 2.5 compared with the performance of PALMER at the same SNR \cite{YW12}. Also, in the localization precision, about a twofold improvement can be made by our analysis \cite{YW12}. Numerical results for various values of $t$ in Fig. \ref{fig2}e-f clearly indicate that the nonuniformity of the background noise does not compromise the performance of our method. 

Finally, let us briefly mention the strengths and the weaknesses of our method. A major strength of our method is the ability to localize multiple molecules with the theoretical limit precision in a high-molecule density. Also, we utilize the separable property of the Gaussian function to derive a set of simple self-consistent equations, which can enhance the fidelity of the multi-emitter localization. On the other hand, the drawback of our approach is that the point spread function requires to be well approximated by the Gaussian function. Also, as a common problem of maximum-likelihood-estimation-based localization methods \cite{AS14}, one needs a prior knowledge about the noise property. 

In summary, we have demonstrated a precise multi-emitter localization analysis that enables the super-resolved imaging at the theoretical-limit speed.  A fast super-resolution microscopy should have an application in, for example, a non-invasive observation of intracellular dynamics at molecular scale. Our method should also provide a powerful means to precisely locate light emitters below the diffraction limit in wide areas of optical science. 
\\
\textbf{Funding.} This work was supported by KAKENHI Grant No. 26287088 from the Japan Society for the Promotion of Science, a Grant-in-Aid for Scientific Research on Innovation Areas ``Topological Materials Science" (KAKENHI Grant No. 15H05855), the Photon Frontier Network Program from MEXT of Japan, and the Mitsubishi Foundation. Y. A. was supported by the Japan Society for the Promotion of Science through Program for Leading Graduate Schools (ALPS).
\\
\textbf{Acknowledgment.} We are grateful for discussions with Y. Okada, K. Goda, H. Mikami and T. Shitara.

\end{document}